\documentclass{article}
\usepackage{fullpage}
\usepackage{mathtools}

\usepackage{amssymb, amsmath, amsfonts, amsthm, graphics, mathrsfs, enumerate}
\usepackage{hyperref}
\hypersetup{
    colorlinks=true,
    linkcolor=blue,
    filecolor=magenta,      
    urlcolor=cyan,
    citecolor = {blue}
    }
\usepackage[hmargin=1 in, vmargin = 1 in]{geometry}
\usepackage{multicol} % multiple columns: \begin{multicols}{2} TEXT HERE \end{multicols}

\usepackage{graphicx}
\usepackage{float}   % let a table or a picture to appear at the desired place
\restylefloat{table} % let a table or a picture to appear at the desired place

\usepackage{algorithm} % algorithm package
\usepackage[noend]{algpseudocode}

\usepackage{natbib}

\DeclareMathOperator*{\argmax}{arg\,max}

\newcommand{\uS}{u^\mathrm{S}}

\newcommand{\sheq}{=}
\newcommand{\ds}{\mathrm{d}s}

\newcommand{\dx}{\mathrm{d}x}

\newcommand{\lpar}{\left(}
\newcommand{\rpar}{\right)}
\newcommand{\Lcb}{\Big\{\hspace*{0.1em}}
\newcommand{\Rcb}{\hspace*{0.1em}\Big\}}
\newcommand{\lsqb}{\left[}
\newcommand{\rsqb}{\right]}
\newcommand{\rabs}{|}
\newcommand{\suite}[1][0ex]{\notag \\[#1] & \mbox{}\hspace{15pt}}
\newcommand{\ig}{\int_\gamma}

\newcommand{\upS}{\upsilon^\mathrm{S}}

\newcommand{\Pbb}{\mathbb{P}}
\newcommand{\Ubb}{\mathbb{U}}
\newcommand{\inv}[1]{\frac{1}{#1}}
\newcommand{\bff}{\boldsymbol{f}}
\newcommand{\bffh}{\hat{\boldsymbol{f}}}
\newcommand{\bftau}{\boldsymbol{\tau}}
\newcommand{\sip}{\cdot}

\title{Optimal Ciliary Locomotion of Axisymmetric Microswimmers}% Force line breaks with \\

\author{Hanliang Guo\footnotemark[1]~\footnotemark[3]~, Hai Zhu\footnotemark[1]~, Ruowen Liu\footnotemark[1]~, Marc Bonnet\footnotemark[2]~, Shravan Veerapaneni\footnotemark[1]~}
\begin{document}
\maketitle
\renewcommand{\thefootnote}{\fnsymbol{footnote}}

\footnotetext[1]{Department of Mathematics, University of Michigan, Ann Arbor, MI, 48109 USA.}
\footnotetext[2]{POEMS (CNRS, INRIA, ENSTA), ENSTA Paris, 91120 Palaiseau, France.}
\footnotetext[3]{Corresponding author. {\em Email address:} {\tt hanliang@umich.edu}}

\begin{abstract}
Many biological microswimmers locomote by periodically beating the densely-packed cilia on their cell surface in a wave-like fashion. While the swimming mechanisms of ciliated microswimmers have been extensively studied both from the analytical and the numerical point of view, the optimization of the ciliary motion of microswimmers has received limited attention, especially for non-spherical shapes. In this paper, using an envelope model for the microswimmer, we numerically optimize the ciliary motion of a ciliate with an arbitrary axisymmetric shape. The forward solutions are found using a fast boundary integral method, and the efficiency sensitivities are derived using an adjoint-based method. Our results show that a prolate microswimmer with a 2:1 aspect ratio shares similar optimal ciliary motion as the spherical microswimmer, yet the swimming efficiency can increase two-fold. More interestingly, the optimal ciliary motion of a concave microswimmer can be qualitatively different from that of the spherical microswimmer, and adding a constraint to the cilia length is found to improve, on average, the efficiency for such swimmers.
\end{abstract}

\section{Introduction}
\label{sc:intro}
%% background
Many swimming microorganisms propel themselves by periodically beating the active slender appendages on the cell surfaces. These slender appendages are known as cilia or flagella depending on their lengths and distribution density.
Eukaryotic flagella, such as the ones in mammalian sperm cells and algae cells, are often found in small numbers, whereas ciliated swimmers such as {\em Paramecium} and {\em Opalina} present more than hundreds of cilia densely packed on the cell surfaces~\citep{Brennen1977, Witman1990}.
Besides the locomotion function for microswimmers, cilia inside mammals serve various other functions such as mucociliary clearance in the airway systems and transport of egg cells in fallopian tubes~(see \citet{Satir2007}, and reference therein). Cilia are also found to be critical in transporting cerebrospinal fluid in the third ventricle of the mouse brain~\citep{Faubel2016} and in creating active flow environments to recruit symbiotic bacteria in a squid-vibrio system~\citep{Nawroth2017}.

Owing to the small length scale of cilia, the typical Reynolds number is close to zero. In this regime, inertia is negligible and the dynamics are dominated by the  viscous effects. As a result, many effective swimming strategies familiar to our everyday life become futile. For example, waving a rigid tail back-and-forth will not generate any net motion over one period. This is known as the time reversibility, or the `scallop theorem', which states that a reciprocal motion cannot generate net motion~\citep{Purcell1977}. Microswimmers therefore need to go through non-time-reversible shape changes to overcome and exploit drag~\citep{Lauga2009Hydrodynamics}.

Ciliated microswimmers break the time-reversibility on two levels. 
On the individual level, each cilium beats in an asymmetric pattern: during the effective stroke, the cilium pushes the fluid perpendicular to the cell surface like a straight rod, and then moves almost parallel to the cell surface in a curly shape during the recovery stroke, in preparation for the next effective stroke.
On the collective level, neighboring cilia beat with a small phase difference that produces traveling waves on the cell surface, namely the metachronal wave.
Existing evidence suggests that the optimal ciliated swimmers exploit the asymmetry on the collective level more than that on the individual level~\citep{Michelin2010Efficiency, Guo2014}.

In this paper, we study the (hydrodynamic) swimming efficiency of ciliated microswimmers of an arbitrary axisymmetric shape. Specifically, the swimming efficiency is understood as the ratio between the `useful power' against the total power. The useful power could be computed as the power needed to drag a rigid body of the same shape as the swimmer with the swim speed while the total power is the rate of energy dissipation through viscous stresses in the flow to produce this motion~\citep{Lighthill1952Squirming}. The goal of this paper is to find the {optimal} ciliary motion that maximizes the swimming efficiency for an arbitrary axisymmetric microswimmer.

Studies of ciliated microswimmers can be loosely classified into two types of models. One type is known as the sublayer models in which the dynamics of each cilium is explicitly modeled, either theoretically~\citep{Brennen1977, Blake1974} or numerically~\citep{Gueron1992, Gueron1993, Guirao2007spontaneous, Osterman2011, Eloy2012kinematics, elgeti2013emergence, Guo2014, ito2019swimming, omori2020swimming}. The other type is known as the {\em envelope model}~{(commonly known as the {\em squirmer model} if the slip profile is time-independent)},  which takes advantage of the densely-packing nature of cilia, and traces the continuous envelope formed by the cilia tips. The envelope model has been extensively applied to study the locomotion of both single and multiple swimmers (e.g., see~\citet{Lighthill1952Squirming, Blake1971spherical, ishikawa2006hydrodynamic, ishikawa2008coherent, Michelin2010Efficiency, vilfan2012optimal, brumley2015metachronal, elgeti2015physics, guo2021optimal, nasouri2021minimum}), as well as the nutrient uptake of microswimmers (e.g.,~\citet{magar2003nutrient, magar2005average, Michelin2011Optimal, michelin2013unsteady}). 
{While originally developed for spherical swimmers, the envelope model has been generalized to spheroidal swimmers (e.g.,~\citet{ishimoto2013squirmer, theers2016modeling}).}

In particular, in a seminal work, \citet{Michelin2010Efficiency} studied the optimal beating stroke for a spherical swimmer using the envelope model. Specifically, the material points on the envelope were assumed to move tangentially on the surface in a time-periodic fashion, hence the swimmer retains the spherical shape. The flow field, power loss, swimming efficiency as well as their sensitivities, thereby, were computed explicitly using spherical harmonics. Their optimization found that the envelope surface deforms in a wave-like fashion, which significantly breaks the time-symmetry at the organism level similar to the metachronal waves observed in biological microswimmers.

Since most biological microswimmers do not have spherical shapes, there is a need for extending the previous work to more general geometries. Such an extension, however, is hard to carry out using semi-analytical methods. Therefore, in this paper, we develop a computational framework for optimizing the ciliary motion of a microswimmer with arbitrary axisymmetric shape. We employ the envelope model, wherein, the envelope is restricted to move tangential to the surface so the shape of the microswimmer is unchanged during the beating period. We use a boundary integral method to solve the forward problem and derive an adjoint-based formulation for solving the optimization problem.  

The paper is organized as follows. In Section~\ref{sc:formulation}, we introduce the optimization problem, derive the sensitivity formulas and discuss our numerical solution procedure. In Section~\ref{sc:results}, we present the optimal unconstrained and constrained solutions for microswimmers of various shape families. Finally, in Section~\ref{sc:conclusion}, we discuss our conclusions and future directions.

\section{Problem Formulation}
\label{sc:formulation}
\subsection{Model} 
\begin{figure}
        \centerline{\includegraphics{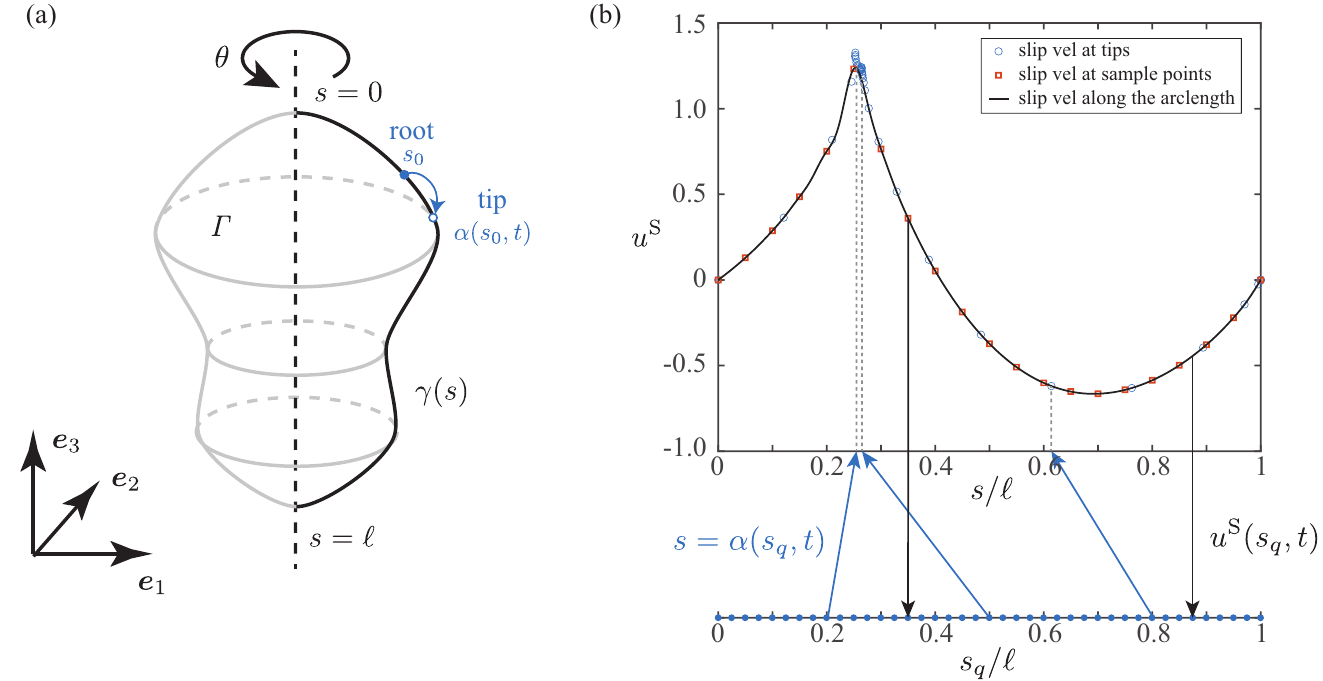}}
\caption[]{(a) Schematic of the microswimmer geometry. The shape is assumed to be axisymmetric, obtained by rotating the generating curve $\gamma$ about the $\boldsymbol{e}_3$ axis. The tip of the cilium rooted at $s_0$ at time $t$ is given by $s=\alpha(s_0,t)$.
(b) Illustration of the algorithm for computing the slip velocity at the quadrature points $u^\mathrm{S}(s_q,t)$. We first compute the ``tip'' position and the corresponding tip velocities (open blue circles) of cilia rooted at the $N_q$ quadrature points $s_q$ (closed blue circles). We then obtain the slip velocities at sample points uniformly distributed along the generating curve (open red squares) by a cubic interpolation. The slip velocity at any arclength (black curve) are then obtained by a high-order B-spline interpolation from the sample points.
We have reduced the number of quadrature and sample points in this figure (compared to values used in the numerical experiments) to avoid visual clutter.
}
	\label{fig:schem}
\end{figure}

Consider an axisymmetric microswimmer whose boundary $\Gamma$ is obtained by rotating a generating curve $\gamma$ of length $\ell$ about $\boldsymbol{e}_3$ axis, as shown in Figure~\ref{fig:schem}(a). 
We adopt the classic envelope model~\citep{Lighthill1952Squirming} and assume that the ciliary tips undergo time-periodic {\em tangential} movements along the generating curve.
Let $s=\alpha(s_0,t)$ be the ciliary tip's arclength coordinate on the generating curve $\gamma$  at time $t$ for a cilium rooted at $s_0$. The tangential slip velocity of this material point in its body-frame is thus
\begin{equation}\label{eq:dispvelo}
u^\mathrm{S}(s, t)= u^\mathrm{S}(\alpha(s_0, t), t)= \partial_t \alpha(s_0, t).
\end{equation}

%---------------------
%------------------------

In addition to the time-periodic condition, the ciliary motion $\alpha$ needs to satisfy two more conditions to avoid singularity~\citep{Michelin2010Efficiency}. 
First, the slip velocities should vanish at the poles 
\begin{equation}\label{eq:Sbc}
\alpha(0,t) = 0 \quad\text{ and } \quad \alpha(\ell,t) = \ell, \quad \forall~ t\in\mathbb{R}^+,
\end{equation}
and second, $\alpha$ should be a monotonic function, that is,
\begin{equation}\label{eq:Sbijective}
\partial_{s_0} \alpha(s_0,t) > 0, \quad \forall~ (s_0,t)\in [0,\ell]\times\mathbb{R}^+.
\end{equation}
The last condition ensures the slip velocity is unique at any arclength $s$; in other words, crossing of cilia is forbidden. While in reality, cilia do cross, this condition is enforced to ensure validity of the continuum model.

In the viscous-dominated regime, the flow dynamics is described by the incompressible Stokes equations at every instance of time
\begin{equation}\label{eq:stokes}
-\mu\nabla^2\boldsymbol{u} + \nabla p = \boldsymbol{0},\quad \nabla\cdot\boldsymbol{u} = 0,
\end{equation}
where $\mu$ is the fluid viscosity, $p$ and $\boldsymbol{u}$ are the fluid pressure and velocity fields respectively.
In the absence of external forces and imposed flow field, the far-field boundary condition  is simply
\begin{equation} \label{eq:far-bc} 
\lim_{\boldsymbol{x}\rightarrow\infty}\boldsymbol{u}(\boldsymbol{x},t) = \boldsymbol0. 
\end{equation}
The free-swimming microswimmer also needs to satisfy the no-net-force and no-net-torque conditions.
Owing to the axisymmetric assumption, the no-net-torque condition is satisfied by construction, and the no-net-force condition is reduced to one scalar equation
\begin{equation}\label{eq:nonetforce}
\int_\Gamma \bff(\boldsymbol{x},t)\cdot\boldsymbol{e}_3\mathrm{d}\Gamma=2\pi\int_\gamma {f}_3(\boldsymbol{x},t)\, x_1\mathrm{d}s= 0,
\end{equation}
where $x_1$ is the $\boldsymbol{e}_1$ component of $\boldsymbol{x}$, $\bff$ is the active force density the swimmer applied to the fluid (negative to fluid traction) and $f_3$ is its $\boldsymbol{e}_3$ component.

Given any ciliary motion $\alpha(s_0,t)$ that satisfies \eqref{eq:Sbc} \& \eqref{eq:Sbijective}, there is a unique tangential slip velocity ${u}^{\mathrm{S}}(s,t)$ defined by \eqref{eq:dispvelo}. 
Such a slip velocity propels the microswimmer at a translational velocity $U(t)$ in the $\boldsymbol{e}_3$ direction, determined by \eqref{eq:nonetforce}. Its angular velocity as well as the translational velocities in the $\boldsymbol{e}_1$ and $\boldsymbol{e}_2$ directions are zero by symmetry. Consequently, the boundary condition on $\gamma$ is given by 
%--------------
\begin{equation}\label{eq:bc}
\boldsymbol{u}(\boldsymbol{x}(s),t) = {u}^{\mathrm{S}}(s,t)\boldsymbol\tau(s)+{U}(t)\boldsymbol{e}_3,
\end{equation}
%--------------
where $\boldsymbol\tau$ is the unit tangent vector on $\gamma$.  
Thereby, the instantaneous power loss $P(t)$ can be written as
\begin{align}
  P(t) &= \int_\Gamma \bff(\boldsymbol{x},t) \cdot \boldsymbol{u}(\boldsymbol{x}, t) \, \mathrm{d}\Gamma \notag\\
  &=  2\pi\left[ \int_\gamma \bff(s,t) \cdot \boldsymbol{\tau}(s) u^\mathrm{S}(s,t)\, x_1 \, \mathrm{d}s + U(t)\int_\gamma \bff(s,t) \cdot \boldsymbol{e}_3\, x_1 \, \mathrm{d}s \right].
\end{align}
The second term on the right-hand-side is zero provided that the no-net-force condition~\eqref{eq:nonetforce} is satisfied.

%%% power loss
Following \citet{Lighthill1952Squirming}, we quantify the performance of the microswimmer by its swimming efficiency $\epsilon$, defined as
\begin{equation}\label{eq:efficiency}
\epsilon = \frac{C_D \langle{U}\rangle^2}{\langle{P}\rangle},
\end{equation}
where ${P}=P(t)$ and ${U}=U(t)$ are the instantaneous power loss and swim speed, $\langle \cdot \rangle$ denotes the time-average over one period, 
and $C_D$ is the drag coefficient defined as the total drag force of towing a rigid body of the same shape at a unit speed along $\boldsymbol{e}_3$ direction. The coefficient $C_D$ depends on the given shape $\gamma$ only; for example, $C_D = 6\pi\mu a$ in the case of a spherical microswimmer with radius $a$.

In our simulations, we normalize the radius of the microswimmer to unity, and the period of the ciliary motion to $2\pi$. It is worth noting that the swimming efficiency~\eqref{eq:efficiency} is size and period independent, thanks to its dimensionless nature. The Reynolds number of a ciliated microswimmer of radius $100\mu \text{m}$ and frequency $30$Hz submerged in water can be estimated as $\mathrm{Re} \sim 10^{-4} $, confirming the applicability of Stokes equations.

\subsection{Numerical algorithm for solving the forward problem}
Before stating the optimization problem, we summarize our numerical solution procedure for the governing equations (\ref{eq:stokes}) -- (\ref{eq:bc}). 
By the quasi-static nature of the Stokes equation~\eqref{eq:stokes}, the flow field $\boldsymbol{u}(\boldsymbol{x},t)$ can be solved independently at any given time, and the time-averages can be found using standard numerical integration techniques (e.g., trapezoidal rule). Here we adopt a boundary integral method (BIM) at every time step.
A similar BIM implementation was detailed in our recent work~\citet{guo2021optimal} which studied the optimization of {\em time-independent} slip profiles. The main procedures are summarized below.

We use the single-layer potential {\em ansatz}, which expresses the velocity as a convolution of an unknown density function $\boldsymbol{\mu}$ with the Green's function for the Stokes equations:
\begin{align}
  \boldsymbol{u}(\boldsymbol{x})
 &= \frac{1}{8\pi}\int_\Gamma \left(\frac{1}{|\boldsymbol{r}|} \mathbf{I} +  \frac{\boldsymbol{r} \otimes \boldsymbol{r} }{|\boldsymbol{r}|^3}\right) \, \boldsymbol{\mu}(\boldsymbol{y}) \, \mathrm{d}\Gamma(\boldsymbol{y}), \quad\text{where}\quad \boldsymbol{r} = \boldsymbol{x} - \boldsymbol{y}.\label{u:SL} \end{align}
 The force density can then be evaluated as a convolution of $\boldsymbol{\mu}$ with the (negative of) traction kernel:
 \begin{align}
  \bff(\boldsymbol{x}) 
 &=\frac{1}{2}\boldsymbol{\mu}\left(\boldsymbol{x}\right) + \frac{3}{4\pi}\int_{\Gamma} \left(\frac{\boldsymbol{r} \otimes \boldsymbol{r} }{|\boldsymbol{r}|^5}\right) (\boldsymbol{r}\cdot\boldsymbol{n}(\boldsymbol{x}))\boldsymbol{\mu}\left(\boldsymbol{y}\right)\mathrm{d}\Gamma\left(\boldsymbol{y}\right). \label{f:SL}
\end{align}
We convert these weakly singular boundary integrals into convolutions on the generating curve $\gamma$ by performing an analytic integration in the orthoradial direction, and apply a high-order quadrature rule designed to handle the $log-$singularity of the resulting kernels~\citep{veerapaneni2009numerical}.
{The Stokes flow problem defined at any time $t$ by equations~\eqref{eq:stokes} --~\eqref{eq:bc} is then recast as the BIM system for the unknowns $\boldsymbol{\mu}$ and $U(t)$ obtained by substituting~\eqref{u:SL} in~\eqref{eq:bc} and~\eqref{f:SL} in~\eqref{eq:nonetforce}. The numerical solution method consists in discretizing $\gamma$ into $N_p$ non-overlapping panels, each panel supporting the nodes of a 10-point Gaussian quadrature rule. The single-layer operator is approximated in Nystr\"om fashion, by collocation at the $N_q=10 N_p$ quadrature nodes, while the values of $\boldsymbol{\mu}$ are sought at the same quadrature nodes.} 
{The resulting BIM system is}
\renewcommand{\arraystretch}{1.5}
\begin{equation}\label{eq:fullsys}
\begin{bmatrix}
\mathcal{S} & -\mathcal{B} \\
 \mathcal{C}& 0
\end{bmatrix}
\begin{bmatrix}
\boldsymbol{\mu}\\
{U(t)}
\end{bmatrix}
=
\begin{bmatrix}
{\boldsymbol{u}^{\mathrm{S}}}\\
{0}
\end{bmatrix},
\end{equation}
%-------
where 
{the vectors $\boldsymbol{\mu}=\boldsymbol{\mu}(s_q,t)$ and $\boldsymbol{u}^{\mathrm{S}}=\boldsymbol{u}^{\mathrm{S}}(s_q,t)$ are the unknown density and the given slip velocity at all quadrature nodes $s_q$,}
$\mathcal{S}$ is the axisymmetric single-layer potential operator (which is fixed for a given shape $\gamma$), $\mathcal{B}$ is the column vector reproducing $\boldsymbol{e}_3$ at each quadrature node, $\mathcal{C}$ is the row vector
such that  $\mathcal{C}[\boldsymbol{\mu}] = \int_\Gamma \bff(\boldsymbol{x})\cdot\boldsymbol{e}_3 \mathrm{d}\Gamma$ is the total traction force in the $\boldsymbol{e}_3$ direction.

The algorithm to obtain the slip velocity at the quadrature nodes at a given time $\boldsymbol{u}^\mathrm{S}(s_q,t)$ is summarized in Figure~\ref{fig:schem}(b). 
Specifically, we start by computing the corresponding ciliary tip position $s=\alpha(s_q,t)$ and the slip velocity  $u^\mathrm{S}(s,t)$ from \eqref{eq:dispvelo}.
These tip positions $s$ can be highly nonuniform, depending on the form of $\alpha$, which could be difficult for the forward solver. 
To circumvent this difficulty and to find a smooth representation of the slip velocities on the quadrature points,
we first find the slip velocities at $N_s$ sample points uniformly distributed along the generating curve by interpolating $u^\mathrm{S}(s,t)$ (we use the routine \texttt{PCHIP} in MATLAB);
the slip velocities at the quadrature nodes $u^\mathrm{S}(s_q,t)$ are then in turn interpolated from the $N_s$ sample points using high-order B-spline bases. 
An alternative approach could be to follow the position and the slip velocity of each material point. In other words, one can use $\boldsymbol{u}^\mathrm{S}(s,t)$ directly on the right-hand-side of \eqref{eq:fullsys}, which will bypass the interpolation steps mentioned above. However, it requires re-assembly of the matrix $\mathcal{S}$ at every time step, significantly increasing the computational cost.

\subsection{Optimization problem}
The goal of this work is to find the optimal ciliary motion for a given arbitrary axisymmetric shape, that is, the ciliary motion $\alpha^{\star}(s_0,t)$ that maximizes the swimming efficiency $\epsilon$:
\begin{equation}\label{eq:unconstrained}
\alpha^{\star}=\argmax_{\alpha\in\mathcal{A}}\epsilon(\alpha),
\end{equation}
where $\mathcal{A}$ is the space of all possible time-periodic ciliary motion satisfying \eqref{eq:Sbc} \& \eqref{eq:Sbijective}.
It is, however, not easy to define and manipulate finite-dimensional parametrizations of $\alpha$ that remain in that space.
To circumvent this difficulty, we follow the ideas in \citet{Michelin2010Efficiency} and represent $\alpha$ in terms of a time-periodic function $\psi(x,t)$, 
such that
\begin{equation}\label{eq:aux}
\alpha(s_0,\psi) = \frac{\ell\int_0^{s_0}{[\psi(x,t)]^2\mathrm{d}x}}{\int_0^{\ell}{[\psi(x,t)]^2\mathrm{d}x}},
\end{equation}
where $\ell$ is the total length of the generating curve $\gamma$.
Note that $\alpha$ is also (implicitly) a function of time $t$, through $\psi = \psi(x,t)$.
It is easy to verify that $\alpha$ given by~\eqref{eq:aux} satisfies the boundary conditions \eqref{eq:Sbc} and the monotonicity requirement \eqref{eq:Sbijective} for any choice of $\psi$.
Conversely, for any $\alpha$ satisfying \eqref{eq:Sbc} and \eqref{eq:Sbijective}, there is at least one $\psi$ that provides $\alpha$. As a result, the optimization problem is recast as finding
\begin{equation}
\psi^{\star} = \argmax_{\psi} \epsilon(\psi), \label{opt:unc}
\end{equation}
where $\psi(\cdot,t)$ is only required to be square-integrable over $[0,\ell]$ for any $t$.

We use a quasi-Newton BFGS method \citep{nocedal2006numerical} to optimize the ciliary motion {\em via} $\psi$, which requires repeated evaluations of efficiency sensitivities with respect to perturbations of $\psi$.
The sensitivities of power loss and swim speed are derived using an adjoint-based method, while the efficiency sensitivity is found using the quotient rule thereafter.
The adjoint-based method exhibits a great advantage against the traditional finite difference method when finding the sensitivities, as regardless of the dimension of the parameter space, the objective derivatives with respect to all design parameters can here be evaluated on the basis of \emph{one} solve of the forward problem for each given ciliary motion $\alpha$.
The derivations are detailed below.

%\vspace{.1in}
\subsection{Sensitivity analysis}
%%%%%%%%%%%%%
% derivations of sensitivities
%%%%%%%%%%%%%
We start by finding the sensitivities in terms of the slip profile $u^\mathrm{S}$. The sensitivities in terms of the auxiliary unknown $\psi$ will be found subsequently by a change of variable. {As the concept of adjoint solution in general rests on duality considerations, we recast the forward flow problem in weak form for the purpose of finding the sought sensitivities of power loss and swim speed, even though the numerical forward solution method used in this work does not directly exploit that weak form. Specifically, we} recast the forward problem \eqref{eq:stokes} -- \eqref{eq:bc} in mixed weak form (see, e.g., \citet[Chap. 6]{brezzi1991mixed}). That is, find $(\boldsymbol{u}, p, \bff, U) \in \boldsymbol{\mathcal{V}}\times\mathcal{P}\times\boldsymbol{\mathcal{F}}\times\mathbb{R},$ such that
\begin{equation}\label{eq:weak}
%\left\{
\begin{array}{lrl}
%%%%%%%%%
\text{(a)} \, & a(\boldsymbol{u}, \boldsymbol{v}) - b(\boldsymbol{v},p) - b(\boldsymbol{u},q) - \langle \bff, \boldsymbol{v}\rangle_{\Gamma} = 0 \hspace{.25in}& \forall (\boldsymbol{v},q) \in \boldsymbol{\mathcal{V}}\times\mathcal{P}\\
%%%%%%%%%
\text{(b)} \, & \langle \boldsymbol{g}, \boldsymbol{e}_3\rangle_\Gamma U + \langle \boldsymbol{g}, u^\mathrm{S} \boldsymbol{\tau} \rangle_\Gamma - \langle \boldsymbol{g}, \boldsymbol{u}\rangle_\Gamma = 0 \hspace{.25in}& \forall \boldsymbol{g} \in \boldsymbol{\mathcal{F}}\\
%%%%%%%%%
\text{(c)} \, & \langle \bff, \boldsymbol{e}_3\rangle_\Gamma = 0\hspace{.25in} & 
\end{array}
%\right.
\end{equation}
where the bilinear forms $a$ and $b$ are defined by
\begin{equation}
a(\boldsymbol{u}, \boldsymbol{v}) := \int_\Omega 2\mu \boldsymbol{D}[\boldsymbol{u}]:\boldsymbol{D}[\boldsymbol{v}]\, \mathrm{d} V, \hspace{.25in} 
b(\boldsymbol{v},q) := \int_\Omega q\, \text{div}\,\boldsymbol{v}\, \mathrm{d}V,
\end{equation}
and $\boldsymbol{D}[\boldsymbol{u}] := (\boldsymbol{\nabla u}+\boldsymbol{\nabla}^{T}\boldsymbol{u})/2$ is the strain rate tensor. $\langle \cdot, \cdot \rangle_\Gamma$ is a short-hand for the inner product on $\Gamma$. For example, $\langle \bff, \boldsymbol{v}\rangle_\Gamma = \int_\Gamma \bff \cdot \boldsymbol{v}\, \mathrm{d}\Gamma$. Similarly, with a slight abuse of notation, the power loss functional could be written as $P(u^\mathrm{S}) := \langle \bff, u^\mathrm{S}\boldsymbol{\tau} + U\boldsymbol{e}_3 \rangle_\Gamma$, where $U := U(u^\mathrm{S})$ is the swim speed functional.

The Dirichlet boundary condition~(\ref{eq:bc}) is (weakly) enforced explicitly through~(\ref{eq:weak}\,b), rather than being embedded in the velocity solution space $\boldsymbol{\mathcal{V}}$, as this will facilitate the derivation of slip derivative identities; 
{this is in fact our motivation for using the mixed weak form~\eqref{eq:weak}.} 
Condition~(\ref{eq:weak}\,c) is the no-net-force condition~\eqref{eq:nonetforce}.

%%%%%%%%%%%%%%%%%%%%%%%

First-order sensitivities of functionals at $u^\mathrm{S}$ are defined as directional derivatives, by considering perturbations of $u^\mathrm{S}$ of the form
\begin{equation}
   u^\mathrm{S}_{\eta} = u^\mathrm{S} + \eta \nu \label{uS:pert}
\end{equation}
for some $\nu$ in the slip velocity space and $\eta\in\mathbb{R}$. Then, the directional (or G\^ateaux) derivative of a functional $J(u^\mathrm{S})$ in the direction $\nu$, denoted by $J'(u^\mathrm{S};\nu)$, is defined as
\begin{equation}
  J'(u^\mathrm{S};\nu) = \lim_{\eta\to0} \frac{1}{\eta}\left( J[u_{\eta}^\mathrm{S}]-J[u^\mathrm{S}] \right). \label{dJ:def}
\end{equation}
For the power loss functional, we obtain (since the derivative of ${u}^\mathrm{S}$ in the above sense is $\nu$)
\begin{equation}
  P'(u^\mathrm{S};\nu) = \langle \bff',{u}^\mathrm{S}\boldsymbol{\tau} + U\boldsymbol{e}_3 \rangle_{\Gamma}   +  \langle \bff,\nu\boldsymbol{\tau} \rangle_{\Gamma} + \langle \bff,\boldsymbol{e}_3 \rangle_{\Gamma}U', \label{dJW:def}
\end{equation}
where $\bff'$ and $U'$ are the derivatives of the active force $\bff$ and swim speed $U$ solving problem~\eqref{eq:weak}, {considered as functionals on the slip velocity $u^\mathrm{S}$}:
\begin{equation}
  \bff' = \lim_{\eta\to0} \frac{1}{\eta}\left( \bff[u_{\eta}^\mathrm{S}]-\bff[u^\mathrm{S}] \right), \qquad
  U' = \lim_{\eta\to0} \frac{1}{\eta}\left( U[u_{\eta}^\mathrm{S}]-U[u^\mathrm{S}] \right).
\end{equation}

Differentiating the weak formulation \eqref{eq:weak} of the forward problem with respect to $u^\mathrm{S}$ leads to the weak formulation of the governing problem for the derivatives $(\boldsymbol{u}', \bff', p', U')$ of the solution $(\boldsymbol{u}, \bff, p, U)$
\begin{equation}
\begin{aligned}
\text{(a) \ }&&	a(\boldsymbol{u}',\boldsymbol{v}) - b(\boldsymbol{u}',q) - b(\boldsymbol{v},p') - \langle \bff',\boldsymbol{v} \rangle_{\Gamma}
 &=0 &\qquad& \forall(\boldsymbol{v},q)\in\boldsymbol{\mathcal{V}}\times\mathcal{P} \\
\text{(b) \ }&&	\langle \nu\boldsymbol{\tau},\boldsymbol{g}\rangle_{\Gamma} + U'\langle \boldsymbol{e}_3,\boldsymbol{g}\rangle_{\Gamma} - \langle \boldsymbol{u}',\boldsymbol{g} \rangle_{\Gamma} &= 0 && \forall\boldsymbol{g}\in\boldsymbol{\mathcal{F}}  \\
  \text{(c) \ }&& \langle \bff',\boldsymbol{e}_3 \rangle_{\Gamma} &= 0
\end{aligned} \label{slip:der:weak}
\end{equation}
Here we have assumed without loss of generality that the test functions in~\eqref{eq:weak} verify $\boldsymbol{v}' = \boldsymbol{0}$, $\boldsymbol{g}' = \boldsymbol{0}$, and $q' = 0$, which is made possible by the absence of boundary constraints in $\boldsymbol{\mathcal{V}}$.

At first glance, evaluating $P'(u^\mathrm{S};\nu)$ in a given perturbation $\nu$ appears to rely on solving the derivative problem~\eqref{slip:der:weak}. However, a more effective approach allows to bypass the actual evaluation of ${\bff}'$. 
Let the adjoint problem be defined by
\begin{equation}
\begin{aligned}
\text{(a) \ }&&	a(\hat{\boldsymbol{u}},\boldsymbol{v}) - b(\hat{\boldsymbol{u}},q) - b(\boldsymbol{v},\hat{p}) - \langle\hat{\bff},\boldsymbol{v}\rangle_{\Gamma}
 &= 0  &\qquad& \forall(\boldsymbol{v},q)\in\boldsymbol{\mathcal{V}}\times\mathcal{P}, \\
\text{(b) \ }&&	\langle \boldsymbol{e}_3,\boldsymbol{g}\rangle_{\Gamma} - \langle \hat{\boldsymbol{u}},\boldsymbol{g}\rangle_{\Gamma} &= 0 && \forall\boldsymbol{g}\in\boldsymbol{\mathcal{F}},
\end{aligned} \label{adj:weak}
\end{equation}
i.e. $(\hat{\boldsymbol{u}},\hat{p})$ are the flow variables induced by prescribing a unit velocity $\boldsymbol{e}_3$ on $\Gamma$. For later convenience, we let $F_0$ denote the (nonzero) net force exerted on $\Gamma$ by the adjoint flow:
\begin{equation}
  F_0 := \langle \hat{\bff},\boldsymbol{e}_3 \rangle_{\Gamma}. \label{F0:def} 
\end{equation}
Problem~\eqref{adj:weak} in strong form is defined by equations~\eqref{eq:stokes} -- \eqref{eq:bc} with $U=1,\,u^\mathrm{S}=0$.
In fact, $F_0$ takes the same value as the drag coefficient $C_D$ in \eqref{eq:efficiency}.

Then, combining the derivative problem~\eqref{slip:der:weak} with the forward problem~\eqref{eq:weak} or the adjoint problem~\eqref{adj:weak} with appropriate choices of test functions allows to derive expressions of $P'(u^\mathrm{S};\nu)$ and $U'(u^\mathrm{S};\nu)$ which do not involve the forward solution derivatives.

Specifically, set the test functions to $(\boldsymbol{v},q,\boldsymbol{g})=(\boldsymbol{u}',p',\bff')$ in equations~(\ref{eq:weak}a,b) of the forward problem and $(\boldsymbol{v},q,\boldsymbol{g})=({\boldsymbol{u}},{p},{\bff})$ in equations~(\ref{slip:der:weak}a,b) of the derivative problem. Then, the combination $(\ref{slip:der:weak}a)+(\ref{slip:der:weak}b)-(\ref{eq:weak}a)-(\ref{eq:weak}b)$ is evaluated, to obtain
\begin{equation}
  \langle \bff',{u}^\mathrm{S}\boldsymbol{\tau} + U\boldsymbol{e}_3 \rangle_{\Gamma}
 = \langle \bff,\nu\boldsymbol{\tau} \rangle_{\Gamma} + \langle \bff,\boldsymbol{e}_3 \rangle_{\Gamma}U'. \label{aux1:alt}
\end{equation}
Substituting \eqref{aux1:alt} into \eqref{dJW:def}, and recalling the no-net-force condition~\eqref{eq:nonetforce}, we have
\begin{equation}
\boxed{
P'(u^\mathrm{S};\nu) = 2\langle \bff,\nu\boldsymbol{\tau} \rangle_{\Gamma}
={4\pi}\int_\gamma  (\bff \cdot \boldsymbol{\tau}) \, \nu x_1 \, \mathrm{d}s.} \label{dJPL:exp:alt}
\end{equation}

Likewise, setting the test functions to $(\boldsymbol{v},q,\boldsymbol{g})=(\boldsymbol{u}',p',\bff')$ in the adjoint problem~\eqref{adj:weak} and $(\boldsymbol{v},q,\boldsymbol{g})=(\hat{\boldsymbol{u}},\hat{p},\hat{\bff})$ in equations~(\ref{slip:der:weak}a,b) of the derivative problem~\eqref{slip:der:weak}, then evaluating $(\ref{slip:der:weak}a)+(\ref{slip:der:weak}b)-(\ref{adj:weak}a)-(\ref{adj:weak}b)$, yields
\begin{equation}
 0 = 
 \langle \hat{\bff},\nu\boldsymbol{\tau} \rangle_{\Gamma}
  + 
 \langle \hat{\bff},U'\boldsymbol{e}_3 \rangle_{\Gamma}
  - 
 \langle \bff',\boldsymbol{e}_3 \rangle_{\Gamma}
 = \langle \hat{\bff},\nu\boldsymbol{\tau} \rangle_{\Gamma} + F_0U'. \label{aux2:alt}
\end{equation}
Note that $\langle \bff',\boldsymbol{e}_3 \rangle_{\Gamma} = 0$ according to (\ref{slip:der:weak}c).
Rearranging terms in \eqref{aux2:alt}, we have
\begin{equation}
\boxed{
  U'(u^\mathrm{S};\nu) = -\frac{1}{F_0}\langle \hat{\bff},\nu\boldsymbol{\tau} \rangle_{\Gamma}
  = - \frac{2\pi}{F_0} \int_\gamma (\hat{\bff}\cdot \boldsymbol{\tau}) \, \nu x_1\, \mathrm{d}s.} \label{dU:exp}
\end{equation}

The sensitivity formulas \eqref{dJPL:exp:alt} \& \eqref{dU:exp}, however, are not practically applicable in this form to the current optimization problem, because the constraints~\eqref{eq:Sbc} \& \eqref{eq:Sbijective} are not easy to enforce on parametrizations of the unknown slip profiles $u^\mathrm{S}$. For this reason, we rewrite the quantities of interest as functionals of $\psi$, and the connection between $\psi$ and $\alpha$ is given by \eqref{eq:aux}.
Specifically, the slip profile is
\begin{equation}
  u^\mathrm{S}(s,t)
 = \partial_{t}\alpha(s_0,\psi)
 = \partial_\psi\alpha(s_0,\psi;\dot{\psi})
 = \partial_\psi\alpha\left( \beta(s,\psi),\psi;\dot{\psi} \right)
 = v^\mathrm{S}(s,\psi), \label{disp:velo:psi}
\end{equation}
where $\dot{\psi} := \partial_t \psi$, and $\beta(s,\psi)$ is the inverse function of $\alpha$, i.e., $s_0 = \beta(s,\psi)$.
The average power loss and swim speed functionals are written as
\begin{equation}
\langle\Pbb\rangle(\psi) := \langle P \rangle(u^\mathrm{S}), \quad \langle \Ubb\rangle(\psi):= \langle U\rangle(u^\mathrm{S}) \qquad
 \text{with \ } u^\mathrm{S}(s,t) = v^\mathrm{S}(s,\psi). \label{Jbb:def}
\end{equation}

On applying the change of variables $s = \alpha(s_0, \psi)$ in the integrals \eqref{dJPL:exp:alt} \& \eqref{dU:exp} and average over one period, we obtain
\begin{align}
\boxed{
\langle\Pbb\rangle'(\psi;\hat{\psi})
 = 2 \int_0^{2\pi}  \int_\gamma \bff(\alpha)\cdot\boldsymbol{\tau}(\alpha)\,x_1(\alpha)\,  v^\mathrm{S}{}'(s,\psi;\hat{\psi}) \,  \partial_s\alpha \, \mathrm{d}s_0 \,\mathrm{d}t,} \label{Jbb'} \\
 \boxed{
  \langle \Ubb\rangle'(\psi;\hat{\psi})
 = -\frac{1}{F_0} \int_0^{2\pi}  \int_\gamma  \hat{\bff}(\alpha)\cdot\boldsymbol{\tau}(\alpha)\,x_1(\alpha)\,v^\mathrm{S}{}'(s,\psi;\hat{\psi}) \,  \partial_s\alpha\, \mathrm{d}s_0 \, \mathrm{d}t,}
\end{align}
where $v^\mathrm{S}{}'(s, \psi; \hat{\psi})$ is the directional derivative  of $u^\mathrm{S}$ with respect to $\psi$ and in the direction $\hat{\psi}$. 
Specifically, we can show that
\begin{multline}\label{eq:vprimeds}
 v^\mathrm{S}{}'(s,\psi;\hat{\psi}) \,  \partial_s\alpha(s_0,\psi) \mathrm{d}s_0
 = \left\{ \partial_s\alpha(s_0,\psi)\,\left[ \partial^2_\psi \alpha \left( s_0,\psi;\hat{\psi},\dot{\psi}\right)
 + \partial_\psi \alpha \left( s_0,\psi;\dot{\hat{\psi}} \right)\,\right] \right. \\
\left. -\partial_{\psi s}\alpha\left( s_0,\psi;\dot{\psi} \right)\;
   \partial_{\psi}\alpha\left( s_0,\psi;\hat{\psi} \right) \right\} \mathrm{d}s_0.
\end{multline}
The derivation and the explicit expression of each term in~\eqref{eq:vprimeds} are given in the Appendix.
%%%%
Finally, the efficiency sensitivity in terms of $\psi$ readily follows by the quotient rule
\begin{equation}\label{eq:effsense}
\epsilon'(\psi;\hat\psi) = C_D\frac{2\langle\Ubb\rangle \langle\Ubb\rangle'\langle\Pbb\rangle - \langle\Ubb\rangle^2\langle\Pbb\rangle'}{\langle\Pbb\rangle^2}.
\end{equation}

%%%%%%%%%%%%%%%%%%%%%%%

%\vspace{.1in}
\subsection{Constraints on surface displacement}
The unconstrained optimization problem~\eqref{opt:unc} introduced above has the tendency to converge to unphysical/unrealistic strokes, where each cilium effectively `covers' the entire generating curve. For a more realistic model, we should add a constraint on the length of the cilium. To this end, and again following \citet{Michelin2010Efficiency}, we {replace the initial unconstrained optimization problem~\eqref{opt:unc} with the penalized optimization problem}
\begin{equation}
\psi^{\star} = \argmax_{\psi} E(\psi), \qquad E(\psi) = \epsilon(\psi) - C(\psi) \label{opt:pen}
\end{equation}
where the (non-negative) {penalty term} $C(\psi)$, defined as
\begin{equation}\label{eq:penalty}
C(\psi) = \int_0^{\ell} H({A}(\psi) - c) \mathrm{d}s_0,
\end{equation}
{serves to incorporate the kinematical constraint $A(\psi)\leq c$ in the optimization problem.} The functional ${A}(\psi)$ in~\eqref{eq:penalty} is a measure of the amplitude of the displacement of individual material points for the stroke (through $\alpha$), and $c$ is a threshold parameter to bound $A(\psi)$ (a smaller $c$ corresponding to a stricter constraint).
$H$ is {a smooth non-negative penalty function defined by}
\begin{equation}
H(u) = \Lambda_1\left[1+\tanh \left(\Lambda_2{u}\right)\right]u^2, \label{H:def}
\end{equation}
{which for large enough $\Lambda_2$ approximates $u\mapsto 2\Lambda_1 u^2 Y(u)$ ($Y$ being the Heaviside unit step function). The multiplicative parameter $\Lambda_1$ then serves to tune the severity of the penalty incurred by violations of the constraint $A(\psi)\leq c$.}
We use $\Lambda_1 = 10^4$ and $\Lambda_2 = 10^{4}$ in our numerical simulations unless otherwise mentioned. The optimization results are not sensitive to the choice of $\Lambda_1$ and $\Lambda_2$. 
A small caveat of the penalty function~\eqref{H:def} is that it has a (small) bump at $\Lambda_2 u \approx -1.109$. This bump would occasionally trap the optimizations into local extrema that have significantly lower efficiencies, depending on the initial guesses. Perturbing $\Lambda_2$ for such cases helps to alleviate the problem.

The physically most relevant definition of ${A}$ would be the actual displacement amplitude of an individual point, i.e., $\Delta s= [\alpha_\text{max}(s_0) - \alpha_\text{min}(s_0)]/2$. The strong nonlinearity of this measure, however, is not appropriate for the computation of the gradient. Following \citet{Michelin2010Efficiency}, we measure the displacement by its variance in time:
\begin{equation}\label{eq:variance}
{A}(\psi) = \langle (\alpha(s_0,\psi) - \langle\alpha\rangle(s_0))^2\rangle.
\end{equation}
The maximum displacement $\Delta s_\text{max} = \max_{s_0}(\Delta s)$ will be found post-optimization for the optimal ciliary motion $\alpha^{\star}$ to better illustrate our results in Section~\ref{sc:results}.

{Like the initial problem~\eqref{opt:unc}, the penalized problem~\eqref{opt:pen} is solvable using unconstrained optimization methods, and we again adopt a quasi-Newton BFGS algorithm} to optimize the ciliary motion. Applying the chain rule to the penalty functional $C(\psi)$, we obtain the derivative of the penalty term in the direction of $\hat\psi$ as
\begin{equation}\label{eq:pensense}
C'(\psi; \hat\psi) = \int_0^\ell H'({A}(\psi) - c) {A}'(\psi; \hat\psi) \mathrm{d}s_0.
\end{equation}
The derivative of the penalized objective functional $E(\psi)$ is therefore 
\begin{equation}
    E'(\psi;\hat{\psi}) = \epsilon'(\psi;\hat{\psi}) - C'(\psi;\hat{\psi}), \label{E':expr}
\end{equation}
where $\epsilon'$ and $C'$ are given by equations \eqref{eq:effsense} and \eqref{eq:pensense}, respectively.

%%%%%%%%%%%%%%%%%%%%%
%%%%%%%%%%%%%%%%%%%%%%%%%%%

\section{Results and discussion}
\label{sc:results}

\subsection{Parametrization}
We parametrize $\psi(s_0,t)$ such that 
\begin{equation}
\psi(s_0,t) = \sum_{k=1}^m \xi_k(t) B_k(s_0),
\end{equation}
where $B_k$ are the 5th order B-spline basis functions and their coordinates $\xi_k(t)$ are expanded as trigonometric polynomials $\xi_k(t) = {a_{0k}}/{2} + \sum_{j=1}^n [a_{jk} \cos jt + b_{jk} \sin jt]$ to ensure time-periodicity. Taken together, we have
\begin{equation}
\psi(s_0,t) = \sum_{k=1}^m \bigg[\frac{a_{0k}}{2}+ \sum_{j=1}^n (a_{jk} \cos jt + b_{jk} \sin jt)\bigg] B_k(s_0)
\label{param:def}
\end{equation}
so that the finite-dimensional optimization problem seeks optimal values for the $m(2n+1)$ coefficients $a_{0k}$, $a_{jk}$ and  $b_{jk}$.
The initial guesses are chosen to be low frequency waves with small wave amplitudes.
To obtain such initial waves, the coefficients of the zeroth Fourier mode $a_{0k}/2$ are randomly chosen from a uniform distribution within $[0,1]$, the first Fourier modes $a_{1k} $ and $b_{1k}$ are randomly chosen from a uniform distribution within $[0,0.01]$, and the coefficients for higher order Fourier modes $j>1$ are set to 0. {To evaluate the gradient of $E(\psi)$ with respect to the design parameters $a_{0k}$, $a_{jk}$ and  $b_{jk}$, we use~\eqref{E':expr} with $\hat{\psi}$ taken as the basis functions of the adopted parametrization~\eqref{param:def}, i.e. $\hat{\psi}(s_0,t)=B_k(s_0)/2$, $\hat{\psi}(s_0,t)=B_k(s_0)\cos jt$ and $\hat{\psi}(s_0,t)=B_k(s_0)\sin jt$, respectively.}
In terms of parametrization, local minima are multiple in the parameter space, since 
multiplying optimal parameters by a constant factor yields the same optimum for $\alpha$.

\subsection{Spheroidal swimmers}
\begin{figure}
        \centerline{\includegraphics{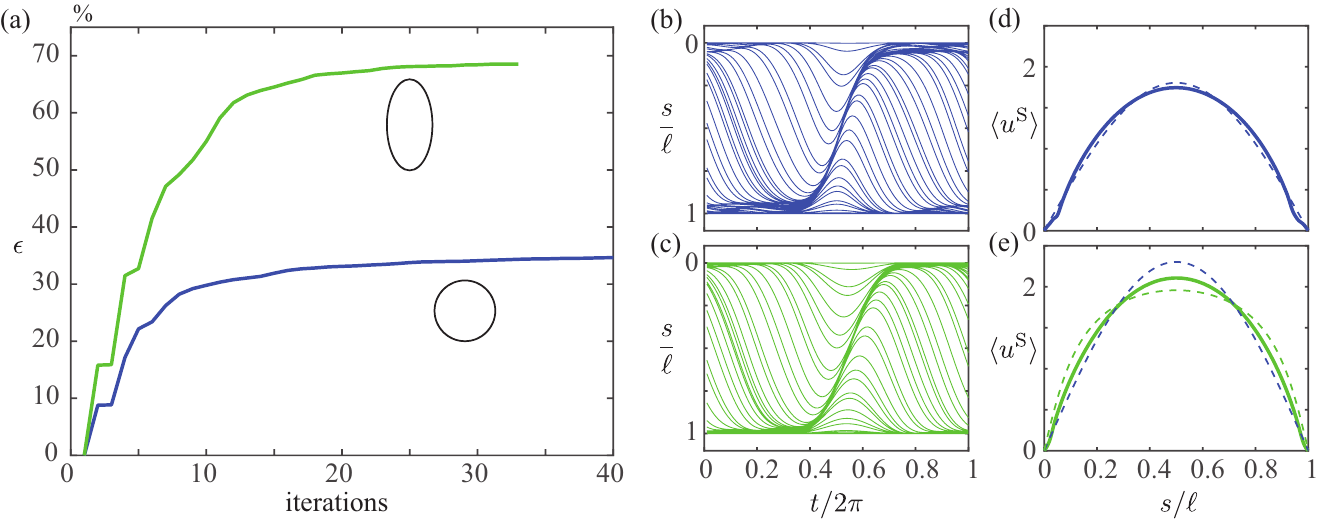}}
\caption[]{Unconstrained optimization history of a spherical swimmer and a prolate swimmer with a 2:1 aspect ratio. The optimal spherical swimmer has an efficiency $\epsilon\approx35\%$ and swim speed $\langle U\rangle\approx1.2$. The optimal prolate swimmer has an efficiency $\epsilon\approx69\%$ and swim speed $\langle U\rangle\approx1.5$.  (a) The efficiency as a function of iterations number. (b) \& (c) The ciliary motions of the optimal swimmers. (d) \& (e) The time-averaged slip velocities (at Eulerian points) are shown in solid curves. Dashed curves are the time-independent optimal slip velocities of the given shape scaled by the swim speed \citep{guo2021optimal}. Parameters used in the optimization: $m = 25, n = 2$. Number of panels $N_p = 20$, number of sample points $N_s = 80$, number of time steps per period $N_t = 50$. Same below unless otherwise mentioned. 
{Note that the vertical axes of figures (b)\&(c) are flipped so that the north pole ($s=0$) appear on the top of the figure. The corresponding waveforms are known as antiplectic metachronal waves (tips are spread out during the effective stroke and close together during the recovery stroke).}
The videos of the optimal ciliary motions can be found in the online supplementary material~(Movie 1~\&~2).}
	\label{fig:optimization_unc}
\end{figure}

By way of validation, we start with optimizing the ciliary motion of a spherical microswimmer. 
The efficiency $\epsilon$ as a function of iteration number for the unconstrained optimization~\eqref{opt:unc} is shown in Figure~\ref{fig:optimization_unc}(a) in blue.  The maximum efficiency is about $35\%$. 
The ciliary motion of the optimal spherical microswimmer is shown in Figure~\ref{fig:optimization_unc}(b). Each curve follows the arclength coordinate of a cilium tip over one period.
We observe, similar to the results of \citet{Michelin2010Efficiency}, clearly distinguished strokes within the beating period. In particular, cilia travel downward `{spread out}' during the effective stroke (corresponding to a stretching of the surface), but travel upward `bundled' together during the recovery stroke in a shock-like structure (corresponding to a compression of the surface).
{This type of waveform is known as an {\em antiplectic metachronal wave}~\citep{Knight-jones1954relations, blake1972model}.}
We note that this optimal ciliary motion produces an efficiency higher than the $23\%$ efficiency obtained numerically by \citet[Fig. 11]{Michelin2010Efficiency}. This is due to a larger maximum displacement $\Delta s_\text{max} \approx 0.45\ell$ in our optimizations (translated to a maximum angle of 81 degrees vs 53 degrees). 
Our optimization result aligns well with their results using the analytical {\em ansatz}~\citep[Fig. 14]{Michelin2010Efficiency}.
Additionally, we found that increasing the number of Fourier mode $n$ increases the maximum displacement as well as the efficiency; the optimal ciliary motion of higher $n$ also exhibits a higher slope for the shock-like structures  (results not shown here). This is again consistent with their analytical {\em ansatz}, which shows that the efficiency approaches $50\%$ in the limit of the maximum displacement approaches 90 degrees, and the corresponding `width' of the shock in this limit is infinitely small.
The mean slip velocity of the Eulerian points within each period are almost identical to the optimal {\em time-independent} slip velocity scaled by the swim speed, as shown in Figure~\ref{fig:optimization_unc}(d).

%---------------------
%------------------------

The optimal unconstrained prolate spheroidal microswimmer with a 2:1 aspect ratio has an efficiency $\epsilon\approx 69\%$, about twice as high as the spherical microswimmer as shown in Figure~\ref{fig:optimization_unc}(a).
This roughly two-fold increase in efficiency is also observed in the optimal time-independent microswimmers~\citep{guo2021optimal}.
The optimal ciliary motion is very close to that of the spherical swimmer (Fig.~\ref{fig:optimization_unc}(b)\&(c)) , while the mean slip velocity of the Eulerian points are between the optimal time-independent slip velocity of the same shape and those of a spherical swimmer, as shown in Figure~\ref{fig:optimization_unc}(e). 
As a sanity check, swapping the ciliary motions obtained from optimizing the spherical swimmer and the prolate swimmer  leads in both cases to lower swimming efficiencies. Specifically, a spherical swimmer with the ciliary motion shown in Figure~\ref{fig:optimization_unc}(c) has 34\% swimming efficiency  and a prolate swimmer with the ciliary motion shown in Figure~\ref{fig:optimization_unc}(b) has 65\% swimming efficiency (compared to 35\% and 69\% using the `true' optimal ciliary motions, respectively).

\begin{figure}
        \centerline{\includegraphics{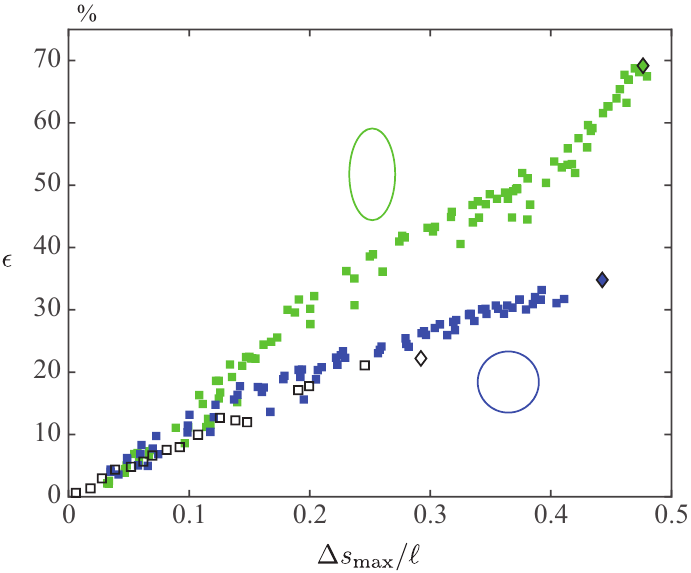}}
\caption[]{Efficiency as a function of maximum displacement of ciliary tips. Blue and green symbols represent spherical and prolate spheroidal swimmers (2:1 aspect ratio) respectively. Diamond symbols are the optimal unconstrained case. Open symbols are optimization results of spherical swimmers taken from \citet[Figure 11]{Michelin2010Efficiency}.
}
	\label{fig:optimization_spheroidal}
\end{figure}

We then turn to the case in which the cilia length is constrained by prescribing a bound on the displacement variance \eqref{eq:variance}.
We control the maximum variance by tuning $c$ in \eqref{eq:penalty}, and the efficiencies are plotted against the maximum displacement $\Delta s_\text{max}$ scaled by the total arclength $\ell$ in Figure~\ref{fig:optimization_spheroidal}. 
Three different random initial guesses are used for each $c$. 
{The unconstrained optimization results for the spherical and prolate spheroidal swimmers are also shown in the figure for reference. Notably, for both the unconstrained swimmers,  the length of the cilia is roughly half the total arclength of the generating curve ($\Delta s_\text{max}\approx \ell/2$). In other words, a cilium rooted at the equator would be able to get very close to both poles during the beating cycle.}
In general, a smaller variance (tighter constraint) leads to a lower efficiency, as expected. The efficiency results of spherical microswimmers closely match those reported by \citet{Michelin2010Efficiency}. 
The efficiencies of the prolate spheroidal microswimmer under constraints are also shown in Figure~\ref{fig:optimization_spheroidal}. Similar to the spherical microswimmer, the efficiency increases roughly linearly with the scaled cilia length $\Delta s_\text{max}/\ell$, and converges to the kinematically unconstrained optimal microswimmer as the maximum variance $c$ is increased.

%---------------------
%------------------------

\begin{figure}
        \centerline{\includegraphics{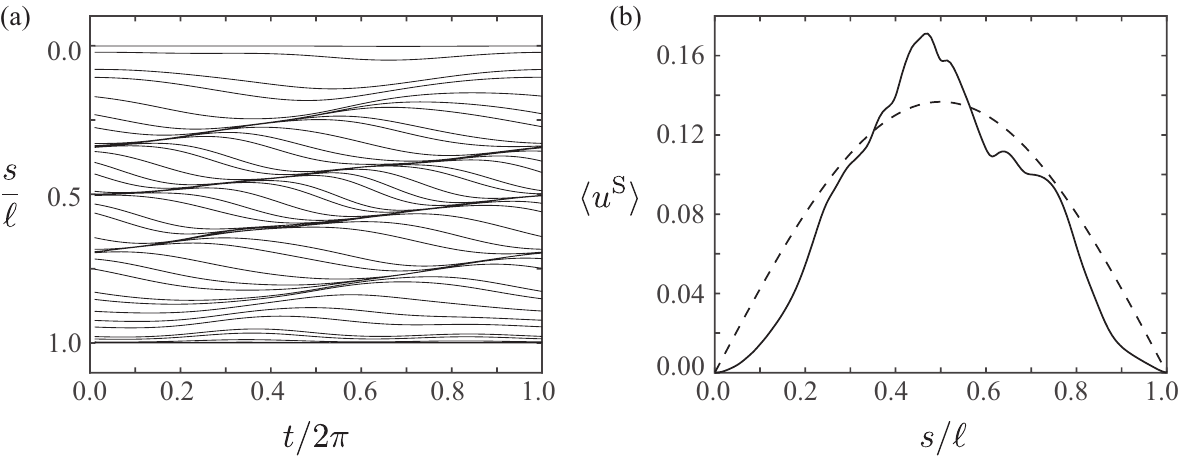}}
\caption[]{Ciliary motion (a) and mean slip velocity (b) for the optimal spherical swimmer with constraint ($\Delta s_\text{max}/\ell \approx 5.0\%$). The efficiency is $\epsilon\approx 6.9\%$, and the swim speed is $\langle U\rangle \approx 0.091$. The swimmer forms multiple waves in the equatorial region, leading to a high slip velocity at $s\approx 0.5\ell$. The motion close to the poles is nearly zero. The dashed curve in (b) is the time-independent optimal slip velocity of the spherical swimmer, scaled by the swim speed.
 The video of the optimal ciliary motion can be  found  in  the  online  supplementary  material~(Movie 3).}
	\label{fig:optimization_spheroidal_waves}
\end{figure}

It is noteworthy that adding a constraint in the cilia length not only limits the wave amplitudes, but also breaks the single wave with larger amplitude into multiple waves with smaller amplitudes~(Fig.~\ref{fig:optimization_spheroidal_waves}(a)), which resemble the metachronal waves of typical ciliated microswimmers such as {\em Paramecium}.
More interestingly, the mean slip velocity in the constrained case can be qualitatively different from the time-independent optimal slip velocity, as shown in Figure~\ref{fig:optimization_spheroidal_waves}(b). In particular, the mean slip velocity around the equator is significantly higher than the time-independent slip velocity, while the mean slip velocity near the poles are closer to zero. This can be inferred from the ciliary motions, as the cilia only move slightly near the poles, whereas multiple waves with significant amplitudes travel around the equator within one period.

%---------------------
%------------------------

\subsection{Non-spheroidal swimmers}\label{sec:concave}
We then investigate the effects of shapes on the optimal ciliary motions and the swimming efficiencies. In particular, we examine whether a single wave travelling between north and south poles always maximizes the swimming efficiency, and whether adding a constraint in the cilia length is always detrimental to the swimming efficiency.

We consider a family of shapes whose generating curves are given by: $(x, z) = (R(\theta)\sin\theta, R(\theta)\cos\theta)$,  where $R(\theta) = (1+\delta\cos 2\theta)$ is a function that makes the radius non-constant, and $\theta\in[0,\pi]$ is the parametric coordinate. For $0<\delta<1$, the radius is the smallest at $\theta = \pi/2$, corresponding to a `neck' around the equator. In the limit $\delta=0$, the generating curve reduces to a semicircle and the swimmer reduces to the spherical swimmer.

%---------------------
\begin{figure}
        \centerline{\includegraphics{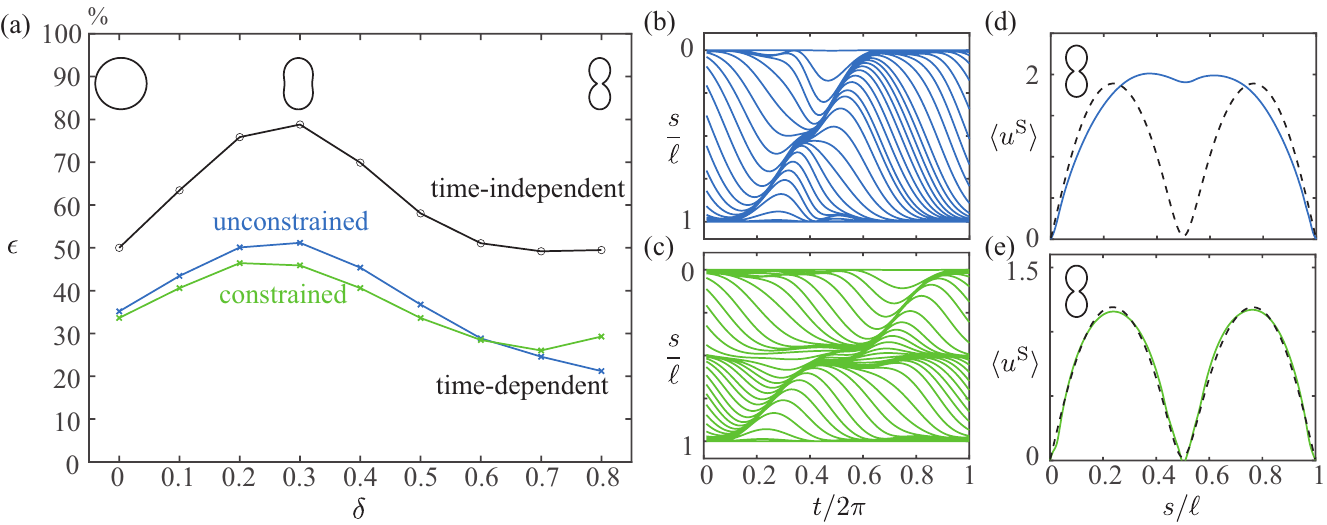}}
\caption[]{
Constrained optimizations could lead to more efficient ciliary motions for microswimmers with a thin `neck' on average.
(a): Efficiencies of the microswimmers with various neck widths. The \textit{median} efficiencies of the time-dependent optimizations across 10 randomized initial conditions are shown for each shape in cross symbols `$\times$'. Unconstrained and constrained optimizations ($c=1$) are depicted in blue and green, respectively.
Efficiencies of the microswimmers with time-independent slips are shown, using black circle symbols `$\circ$', as a reference.
(b)\&(c): Ciliary motions of microswimmers with $\delta=0.8$ from unconstrained and constrained optimizations from the same initial guess. The swimming efficiencies are $20\%$ and $29\%$, respectively.
(d)\&(e): Mean slip velocity corresponding to the ciliary motions in (b)\&(c). Blue dashed curves are the optimal time-independent slip velocities scaled by the swim speed.
In these simulations, we increase the number of panels $N_p = 40$ to resolve the sharp shape change.
The videos of the optimal ciliary motions can be  found  in  the  online  supplementary  material~(Movie 4~\&~5)}
	\label{fig:optimization_neck}
\end{figure}
%------------------------

The optimization results are depicted in Figure~\ref{fig:optimization_neck} for $0\le \delta \le 0.8$. Some corresponding shapes are shown as insets. 
The median efficiencies of ten Monte Carlo simulations are plotted for each $\delta$ value, and compared against the time-independent efficiencies.
For all three cases (constrained, unconstrained, and time-independent), the efficiencies increase as $\delta$ increases from $0$ to $0.3$. This is because increasing $\delta$ in this regime makes the shape more elongated.
Increasing $\delta$ further reduces the efficiencies as the `neck' at the equator becomes more and more pronounced. 
Additionally, the unconstrained microswimmers, on average, have better efficiencies than the microswimmers with kinematic-constraints for $0\le \delta\le 0.6$.

Interestingly, unconstrained optimization may result in worse ciliary motions on average when the shape is highly curved, compared to its kinematically-constrained counterpart. 
Specifically, the constrained microswimmers have higher median efficiencies for $\delta \ge 0.7$.
We note that the unconstrained optimizations are likely to be trapped in local optima where the ciliary motion forms a single wave (Fig.~\ref{fig:optimization_neck}(b)), whereas the constrained optimizations are `forced' to find the ciliary motion with multiple waves split at the equator (Fig.~\ref{fig:optimization_neck}(c)), because of the constrained cilia length. 
Additionally, our numerical results show that a single wave travelling between the north and south poles is not as efficient as two separate waves travelling within each hemisphere for this shape. Figures~\ref{fig:optimization_neck}(d)\&(e) show that the single wave generates a high mean slip velocity at the position where the generating curve bends inward (the equator), whereas the two separate waves generate a mean slip velocity similar to that obtained from the time-independent optimization.
In a way, the constraint in cilia length is helping the optimizer to navigate the parameter space.

\begin{figure}
        \centerline{\includegraphics{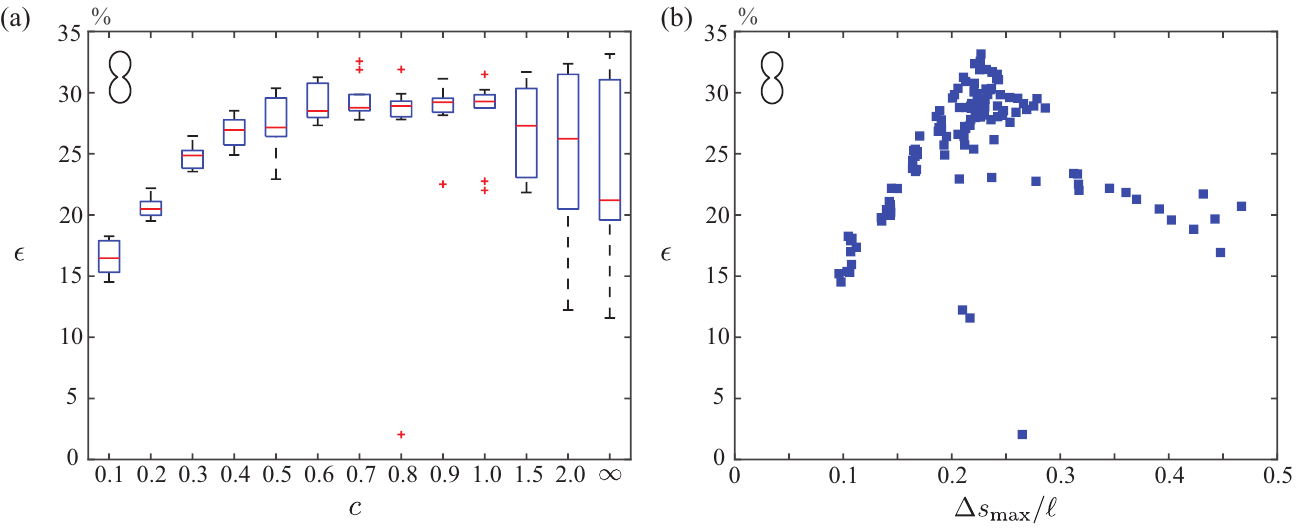}}
\caption[]{
Statistical results of thin neck microswimmer of $\delta = 0.8$ with various constraint $c$ for 10 Monte-Carlo simulations. The unconstrained simulation is denoted by $c=\infty$.
(a) Efficiencies grouped by the constraint $c$. For each box, the central mark indicates the median of the 10 random simulations, and the bottom and top edges of the box indicate the 25th and 75th percentiles, respectively. The outliers are denoted by red $+$ symbols.
(b) Efficiencies plotted against the maximum displacement $\Delta s_\text{max}/\ell$.
The numerical parameter $\Lambda_2$ is set to be $10^4$ by default. Occasionally the optimization might stop within merely a few iterations, making the ciliary motion stuck in a very inefficient local minimum. Setting $\Lambda_2$ to $10^3$ for these cases (most of the time) cures the problem.
}
	\label{fig:optimization_neck_box}
\end{figure}

To better understand the effects of constraints on the highly curved shapes, we present the statistical results of the thin neck microswimmer ($\delta = 0.8$) with various constraints in Figure~\ref{fig:optimization_neck_box}.
In general, the highest efficiency from  the Monte Carlo simulations increases with the constraint for $c\le0.8$, similar to the case of spheroidal swimmers (Figure~\ref{fig:optimization_spheroidal}). 
Keep increasing $c$ has limited effect on the highest efficiencies, indicating that the constraint is no longer limiting the optimal ciliary motion.
The median efficiencies (red horizontal lines), on the other hand, decreases with the constraint if $c \ge 0.8$, consistent with the observation from Figure~\ref{fig:optimization_neck}.
It is worth noting that the constrained optimization is more likely to get stuck in {\em very} low efficiencies (e.g., the lowest outlier for $c=0.8$), possibly due to the secondary bump of the penalty function $C$ mentioned earlier.

%---------------------
%------------------------

All data points from the optimization are plotted in Figure~\ref{fig:optimization_neck_box}(b) as function of the maximum displacement $\Delta s_\text{max}$.
The efficiencies grow almost linearly until $\Delta s_\text{max} \approx 0.25 \ell$, as in the case of spheroidal swimmers, and decrease for larger $\Delta s_\text{max}$. This is another evidence that the optimal ciliary motion for this shape consists of two separate waves traveling within each hemisphere. 
We want to emphasize that unconstrained optimization can still reach the optimal ciliary motion, as shown in the box of $c=\infty$. However it is more likely to reach the sub-optimal ciliary motion compared to the constrained cases.

\section{Conclusions and Discussions}
\label{sc:conclusion}
In this paper, we extended the work of \citet{Michelin2010Efficiency} and studied the optimal ciliary motion for a microswimmer with arbitrary axisymmetric shape.
In particular, the forward problem is solved using a boundary integral method and the sensitivities are derived using an adjoint-based method.
The auxiliary function $\psi$ is parametrized using high-order B-spline basis functions in space and a trigonometric polynomial in time.
We studied the constrained and unconstrained optimal ciliary motions of microswimmers with a variety of shapes, including spherical, prolate spheroidal, and concave shapes which are narrow around the equator.
In all cases, the optimal swimmer displays (one or multiple) traveling waves, reminiscent of the typical metachronal waves observed in ciliated microswimmers.
{Specifically, for the spherical swimmer with limited cilia length (Fig.~\ref{fig:optimization_spheroidal_waves}(a)), the ratio between the metachronal wavelength close to the equator and the cilia length could be estimated as ${\lambda_{MW}}/{\Delta s_\text{max}}\approx{0.2\ell}/{0.05\ell} = 4$. This ratio lies in the higher end of the data collected in~\citet[Table 9]{rodrigues2021bank} for biological ciliates, which reports ratio ranging between $0.5$ to $4$. Our slightly high ratio estimate may not be surprising after all, as the envelope model prohibits the crossing between neighboring cilia.}

We showed that the optimal ciliary motions of prolate microswimmer with a 2:1 aspect ratio are very close to the ones of spherical microswimmer, while the swimming efficiency can increase two-fold. 
The mean slip velocity of unconstrained microswimmers also tend to follow the optimal time-independent slip velocity, which can be easily computed using our recent work~\citep{guo2021optimal}.

Most interestingly, we found that constraining the cilia length for some shapes may lead to a better efficiency on average, compared to the unconstrained optimization. It is our conjecture that this counter-intuitive result is because the constraint effectively reduces the size of the parameter space, hence lowering the probability of being trapped in local optima during the optimization.
{Although the concave shapes studied in Section~\ref{sec:concave} are somewhat non-standard, they allows us to gain insights into the effect of local curvature on optimal waveform. Incidentally, these shapes are also observed for ciliates in nature (e.g. during the cell division process).}

{It is worth pointing out that works on sublayer models (explicitly modeling individual cilia motions) have reported swimming or transport efficiencies in the orders of $0.1\sim1\%$ (see, e.g., \citet{elgeti2013emergence, ito2019swimming, omori2020swimming}), {\em much} lower than the optimal efficiency reported here and others using the envelope models.
This large difference can possibly be attributed to the fact that the envelope model we adopted here considers only the energy dissipation {\em outside} the ciliary layer (into the ambient fluid), while sublayer models in general considers energy dissipation both inside and outside the ciliary layer. Research has shown that the energy dissipation inside the layer could be as high as $90\sim95\%$ of the total energy dissipation, due to the large shear rate inside the layer (see, e.g., \citet{keller1977porous, ito2019swimming}). We note that it is possible to incorporate energy dissipation {\em inside} the ciliary layer in the envelope model, as previously done in \citet{vilfan2012optimal}, albeit for a time-independent slip profile. Additionally, the difference could also be due to modeling assumptions on the cilia length and the number of cilia. In particular, the cilia length considered in sublayer models are usually below $1/10$ of the body length. \citet{omori2020swimming} showed that the swimming efficiency increases with the cilia length as fast as powers of 3 in the short cilia limit, and the number of cilia also has a significant positive effect on the swimming efficiency (the envelope model assumes a ciliary continuum). Factoring all three factors (energy inside/outside, cilia length, number of cilia) could bridge the gap between the results obtained from these two types of models.}

It is without a doubt that maximizing the hydrodynamic swimming efficiency is not the sole objective for biological microswimmers. Other functions such as generating feeding currents~\citep{Riisgaard2010, Pepper2013} and creating flow environment to accelerate mixing for chemical sensing~\citep{Supatto2008, Shields2010, Ding2014, Nawroth2017} are also important factors to consider as a microswimmer. The effect of such multi-tasking on the ciliary dynamics is not well understood. Nevertheless, our work provides an efficient framework to investigate the hydrodynamically optimal ciliary motions for microswimmers of any axisymmetric shape, and could provide insights into designing artificial microswimmers. 

A straightforward extension of our work is to allow more general ciliary motions, e.g., including deformations normal to the surface. Such a swimmer will display time-periodic shape changes and the optimization will require the derivation of shape sensitivities. Additionally, the computational cost would also increase significantly because the matrix in~\eqref{eq:fullsys} needs to be updated at every time step.
Our framework is also open to many generalizations and could for example help in accounting for the multiple factors mentioned above, such as mixing for chemical sensing, in the study of optimal ciliary dynamics.

\vspace{.2in}
\noindent
{\bf Acknowledgments.}
Authors gratefully acknowledge support from NSF under grants DMS-1719834,  DMS-1454010 and DMS-2012424.

%\backsection[Supplementary data]{\label{SupMat}Supplementary movies are available at \\https://doi.org/10.1017/xxxx}
%
%\backsection[Funding]{Authors gratefully acknowledge support from NSF under grants DMS-1719834,  DMS-1454010 and DMS-2012424.}
%
%\backsection[Declaration of interests]{The authors report no conflict of interest.}

\appendix
%\counterwithin{figure}{section}
\section*{Appendix A: Derivations of sensitivities}
In this Appendix, we include the detail derivations that lead to \eqref{eq:vprimeds} and the explicit expressions of the terms therein.

Recall that the power loss and the swim speed can be written as functionals of $\psi$, as shown in \eqref{Jbb:def}.
The sensitivities of $\langle \Pbb\rangle$ and $\langle \Ubb\rangle$ can thus be formulated by considering perturbed versions of $\psi$ as in
\begin{equation}
  \psi_{\eta}(x,t)
 = \psi(x,t) + \eta \hat{\psi}(x,t), \qquad \text{i.e. \ } \psi_{\eta} = \psi + \eta \hat{\psi}, \label{psi:pert}
\end{equation}
so that the perturbed location $s_{\eta}$ at time $t$ of the material particle initially located at $s_0$ is given by
\begin{equation}
  s_{\eta} = \alpha(s_0,\psi_{\eta}),
\end{equation}
the functional $\alpha$ being unchanged. 
Similar to \eqref{disp:velo:psi}, the perturbed slip velocity $\uS_{\eta}(s,t)$ satisfies
\begin{equation}
  \uS_{\eta}(s,t)
 = \partial_\psi\alpha\lpar \beta(s,\psi_{\eta}),\psi_{\eta};\dot{\psi}_{\eta} \rpar = \upS(s,\psi_{\eta}), \label{disp:velo:eta}
\end{equation}
where $\beta$, the inverse function of $\alpha$, is also unchanged.

Notice that $\uS$ and $\uS_{\eta}$ given by~\eqref{disp:velo:psi} and~\eqref{disp:velo:eta} are evaluated at the same time $t$ and current location $s$ (the latter being thus reached from different initial positions $\beta(s,\psi)$ and $\beta(s,\psi_{\eta})$). 
This allows us to define the directional derivative $\upS{}'(s,\psi;\hat{\psi})$ of $\uS$ with respect to $\psi$ in the direction $\hat{\psi}$, as a total derivative with respect to $\eta$:
\begin{equation}
  \upS{}'(s,\psi;\hat{\psi}) := \lim_{\eta\to0} \inv{\eta} \lsqb \uS_{\eta}(s,t) - \uS(s,t) \rsqb
 = \frac{\text{d}}{\text{d}\eta} \partial_\psi\alpha \left.\lpar \beta(s,\psi_{\eta}),\psi_{\eta};\dot{\psi}_{\eta} \rpar \right\rabs_{\eta=0}
\end{equation}
Carrying out the above differentiation in a straightforward way, we find
\begin{multline}
   \upS{}'(s,\psi;\hat{\psi})
 = \partial_{\psi s}\alpha\lpar \beta(s,\psi),\psi;\dot{\psi} \rpar\;
   \partial_{\psi}\beta\lpar s,\psi;\hat{\psi} \rpar \\
 + \partial_{\psi\psi}\alpha\lpar \beta(s,\psi),\psi;\dot{\psi}\,,\, \hat{\psi} \rpar
 + \partial_\psi\alpha\lpar \beta(s,\psi),\psi;\dot{\hat{\psi}} \rpar. \label{uS':aux}
\end{multline}
Moreover, for any $\psi$, the functions $\alpha$ and $\beta$ are linked through
\begin{equation}
  s = \alpha\lpar \beta(s,\psi),\psi \rpar
\end{equation}
which, upon taking the directional derivative in the direction $\hat{\psi}$ and using the chain rule, yields
\begin{equation}
  0 = \partial_s\alpha\lpar \beta(s,\psi),\psi \rpar
  \partial_{\psi}\beta\lpar s,\psi;\hat{\psi} \rpar
  + \partial_{\psi}\alpha\lpar \beta(s,\psi),\psi;\hat{\psi} \rpar.
\end{equation}
The above equality allows us to eliminate $\partial_{\psi}\beta$ from~\eqref{uS':aux}, to obtain
\begin{align}
   \upS{}'(s,\psi;\hat{\psi})
 &= -\partial_{\psi s}\alpha\lpar \beta(s,\psi),\psi;\dot{\psi} \rpar\;
   \frac{\partial_{\psi}\alpha\lpar \beta(s,\psi),\psi;\hat{\psi} \rpar}
        {\partial_s\alpha\lpar \beta(s,\psi),\psi \rpar} \suite
 + \partial_{\psi\psi}\alpha\lpar \beta(s,\psi),\psi\;;\;\dot{\psi}\,,\, \hat{\psi} \rpar
 + \partial_\psi\alpha\lpar \beta(s,\psi),\psi;\dot{\hat{\psi}} \rpar.  \label{der:upS}
\end{align}

In practice, the slip velocity derivative $\upS{}'$ given by~\eqref{der:upS} is more conveniently expressed in the initial arclength variable $s_0=\beta(s,\psi)$. Moreover, in the event that $\psi(s_0,t)\sheq0$ for some $s_0$ and $t$, $\upS{}'$ given by~\eqref{der:upS} blows up since $\partial_s \alpha(\beta(s,\psi),\psi) = 0$ in this case, whereas $\upS{}'\ds$ remains finite if expressed in terms of $s_0$ (since $\ds = \partial_s\alpha(s_0,\psi) \ds_0$). Upon effecting the change of variable $s=\alpha(s_0,\psi)$ in the integrals~\eqref{dJPL:exp:alt} and~\eqref{dU:exp}, we obtain
\begin{align}
\langle\Pbb\rangle'(\psi;\hat{\psi})
&= 4\pi \left\langle\ig R(\alpha(s_0,\psi))\, \bff(\alpha(s_0,\psi),t)\sip\bftau(\alpha(s_0,\psi))\,\upS{}'(s,\psi;\hat{\psi}) \,  \partial_s\alpha(s_0,\psi) \ds_0 \right\rangle\\
\langle\Ubb\rangle'(\psi;\hat{\psi})
&= \frac{-2\pi}{F_0}\left\langle \ig R(\alpha(s_0,\psi))\, \bffh(\alpha(s_0,\psi))\sip\bftau(\alpha(s_0,\psi))\,\upS{}'(s,\psi;\hat{\psi}) \,  \partial_s\alpha(s_0,\psi) \ds_0 \right\rangle
\end{align}
where, thanks to~\eqref{der:upS}, we have used
\begin{align}
  \upS{}'(s,\psi;\hat{\psi}) \ds
 &= \upS{}'(s,\psi;\hat{\psi}) \,  \partial_s\alpha(s_0,\psi) \ds_0 \notag\\
 &= \Lcb \partial_s\alpha(s_0,\psi)\,\lsqb \partial^2_\psi \alpha \lpar s_0,\psi;\hat{\psi},\dot{\psi}\rpar
 + \partial_\psi \alpha \lpar s_0,\psi;\dot{\hat{\psi}} \rpar\,\rsqb \suite\qquad
    -\partial_{\psi s}\alpha\lpar s_0,\psi;\dot{\psi} \rpar\;
   \partial_{\psi}\alpha\lpar s_0,\psi;\hat{\psi} \rpar \Rcb \ds_0.
\end{align}
This completes our derivation of \eqref{eq:vprimeds}.

%%%%
For the ciliary motion~\eqref{eq:aux} used here, introducing the shorthand notation $I(f,g;s):=\int_0^s f(x)g(x) \dx$, we have
\begin{align}
  \alpha(s_0,\psi)
 &= \frac{\ell I(\psi,\psi;s_0)}{I(\psi,\psi;\ell)} \\
  \partial_s\alpha(s_0,\psi)
 &= \frac{\ell\psi^2(s_0)}{I(\psi,\psi;\ell)} \\
  \partial_{\psi}\alpha\lpar s_0,\psi;\hat{\psi} \rpar
 &= \frac{2\ell I(\psi,\hat{\psi};s_0)}{I(\psi,\psi;\ell)}
  - 2\alpha(s_0,\psi)\frac{I(\psi,\hat{\psi};\ell)}{I(\psi,\psi;\ell)} \\
  \partial_{s\psi}\alpha\lpar s_0,\psi;\dot{\psi} \rpar
 &= \frac{2\ell\psi(s_0)\dot{\psi}(s_0)}{I(\psi,\psi;\ell)}
 - 2\ell\frac{I(\psi,\dot{\psi};\ell)\, \psi^2(s_0)}{\lpar I(\psi,\psi;\ell) \rpar^2} \\
  \partial^2_\psi \alpha \lpar s_0,\psi\;;\;\hat{\psi},\dot{\psi} \rpar
 &= \frac{2\ell I(\hat{\psi},\dot{\psi};s_0)}{I(\psi,\psi;\ell)}
 -2\alpha(s_0, \psi) \frac{I(\hat{\psi},\dot{\psi};\ell)}{I(\psi,\psi;\ell)} \suite
 -\frac{2I(\psi,\hat{\psi};\ell)}{I(\psi,\psi;\ell)}
  \partial_{\psi}\alpha\lpar s_0,\psi;\dot{\psi} \rpar
 -\frac{2I(\psi,\dot{\psi};\ell)}{I(\psi,\psi;\ell)}
  \partial_{\psi}\alpha\lpar s_0,\psi;\hat{\psi} \rpar.
\end{align}

%%%%%%%%%%%%%%%%%%%%%%%%%%
\section*{Appendix B: Initial coefficient sensitivity}
{In our optimizations, the initial guesses are chosen to be low-frequency waves with small wave amplitudes. This is obtained by choosing the coefficients of the first Fourier modes from a uniform distribution within $[0,0.01]$ (to restrict the initial wave amplitudes), and setting the coefficients of the higher modes to 0 (to discourage high-frequency waves).}

{Restricting our attention to low-frequency waves effectively sets a time scale in our problem. That is, it helps us to focus on the ciliary motion within {\em one} beating cycle which is given by the base Fourier mode. Note that there is a danger of confusing the (spatial) Legendre modes used in \citet{Blake1971spherical} and the (temporal) Fourier modes studied here. While the swim speed is determined by the first Legendre mode, introducing higher order Fourier modes would affect the swim speed. Specifically, cilia beating twice as fast (beating two cycles in the same time span) could double the swim speed. However, the efficiency would remain unchanged because of the simultaneous increase of the power loss.}

\begin{figure}
        \centerline{\includegraphics{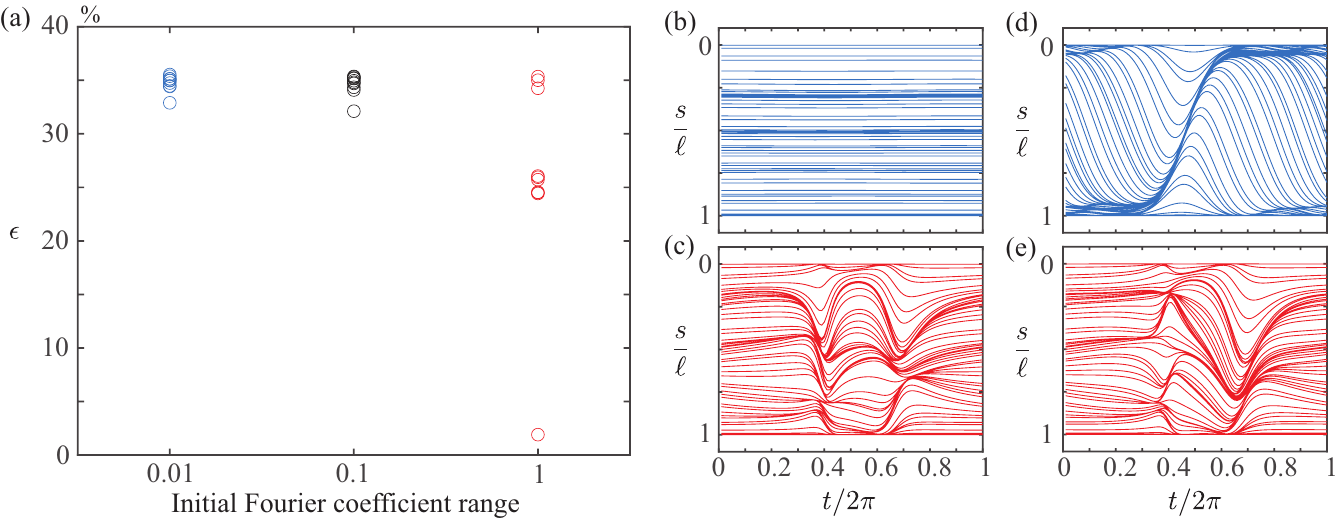}}
\caption[]{Sensitivity to the initial Fourier coefficient. 
		(a) Optimized efficiencies for the unconstrained spherical swimmer with the initial first Fourier mode chosen from $[0,0.01]$, $[0,0.1]$, $[0,1]$ respectively. 
		(b)\&(d) The initial and final waveforms of the case where the range is $[0,0.01]$.
		(c)\&(e) The initial and final waveforms of the case where the range is $[0,1]$.}
	\label{fig:initialcoef}
\end{figure}

{Due to the high-dimensional nature of the problem (hundreds of degrees of freedom), many local optima exist. As shown in Figure~\ref{fig:initialcoef}(a), a large initial range of the Fourier coefficient (e.g., $[0,1]$) increases the risk of the optimizer getting stuck close to an unsuitable local optimum. For example, an initial waveform as shown in Figure~\ref{fig:initialcoef}(c) can only be optimized to a waveform shown in Figure~\ref{fig:initialcoef}(e), which has a swimming efficiency as low as $2\%$. On the other hand, the initial wave with small amplitudes (as shown in Figure~\ref{fig:initialcoef}(b)) could almost always be optimized to the waveform with swimming efficiency $\epsilon\approx35\%$.}

%---------------------
%------------------------

\bibliographystyle{unsrtnat}
\bibliography{references}% Produces the bibliography via BibTeX.

\end{document}